\documentclass{emulateapj}
\usepackage{colordvi}
\usepackage{natbib}

\newcommand{\msun}{$M_{\odot}$}

\newcommand{\fesc}{$f_{esc}$}

\def\f0{$F_0$}

\input epsf

\begin{document}
\title{Local Group Dwarf Galaxies and the Contribution of the First Stars 
to Reionization} 
\author{JASON TUMLINSON}\affil{Yale Center for Astronomy and Astrophysics, 
Departments of Physics and Astronomy, 
Yale University, P. O. Box 208121, New Haven, CT 06520}
\begin{abstract}
The Local Group contains some dwarf galaxies that apparently formed all
their stars early, lost their gas to winds or ionization, and survive
today as ``fossils'' from the epoch of reionization. This study presents
new models of these objects based on the hierarchical chemical evolution
framework of Tumlinson (2006). The model accurately reproduces the
observed luminosity-metallicity relation of fossil dwarf galaxies with
minimal free parameters. When calibrated to this relation, the models show
that small dark matter ``minihalos'' formed $2 - 8$\% of their baryonic
mass into stars prior during reionization.  By tracking the chemical
enrichment of these early halos the models specify that metal-free first
stars contributed approximately $5 - 10$\% of the ionizing photons
generated by these small halos and so did not dominate reionization. 
Models that allow for larger relative contributions from metal-free stars
may not generate enough total ionizing photons for early reionization.
As significant star formation in early minihalos is a requirement of
many successful models for the IGM reionization, these models can be
considered to pass a key test of their validity. It appears we have as
much to learn about reionization from the universe at $z = 0$ as at $z =
6$. \end{abstract} 

\section{Introduction}

The reionization of the intergalactic medium was a major change in the
physical state of most of the baryons in the Universe, potentially a
major influence on the formation and development of galaxies, and may
be intimately connected with the first generations of metal-free stars.
Observations by the Wilkinson Microwave Anisotropy Probe ({\em WMAP}) of
the integrated Thomson scattering optical depth to the cosmic microwave
background, $\tau_{es} = 0.17^{+0.08}_{-0.07}$, and the Gunn-Peterson
absorption trough in spectra of high-$z$ QSOs found by the Sloan Digital
Sky Survey (SDSS), together suggest that reionization had a complicated
history in the interval $z = 6 - 20$, which spans $< 1$ Gyr of cosmic
time\footnote{I assume the cosmological parameters derived from first-year
WMAP data by Spergel et al. (2003); $\Omega_m = 0.27$, $\Omega_{\Lambda}
= 0.73$, $\Omega_b = 0.044$, and $H_0 = 71$ km s$^{-1}$ Mpc$^{-1}$.}.

Theoretical studies have attempted to explain these data in terms of
semi-analytic reionization models that calculate the collapse of dark
matter halos and track the growth of cosmological \ion{H}{2} regions
(Venkatesan, Tumlinson, \& Shull 2003; Haiman \& Holder 2003; Cen 2003;
Tumlinson, Venkatesan, \& Shull 2004, hereafter TVS04).  These studies
vary in their particulars, but they generally include assumptions about
or {\em ab initio} calculations of three important physical quantities
that have so far been unconstrained by observation.  The first of these is
$f_*$, the efficiency of converting virialized baryons within dark-matter
halos into stars. This factor may depend on such local conditions as gas
density, metallicity, and radiation feedback in unknown fashion, and
as $f_*$ is difficult to calculate a priori, it is generally assumed.
The second uncertain parameter is $\gamma _0$, the time-integrated
ionizing photon production per baryon in stars. This quantity is
essentially completely determined by the initial mass function (IMF) at
a given metallicity, and ranges from $\gamma _0$ = 4000 for a Salpeter
IMF at solar metallicity to $\sim 50000 - 100000$ for an IMF dominated
by massive stars, as expected at low metallicity. TVS04 used detailed
chemical abundances in metal-poor Galactic halo stars have favored
massive star IMFs with $\gamma \simeq 50000 - 80000$ for metal-free
stars.  Finally, the escape fraction of ionizing photons from their
parent galaxies is denoted by \fesc\ and has been estimated at 0.1 for
bright galaxies in the local universe, but is unconstrained for small
galaxies in the high-redshift universe. Thus, the three critical model
inputs for reionization are all uncertain.

Theoretical studies attempting to explain the reionization data have
varied in their specific treatment of these uncertain parameters,
specifically with regard to their assumptions about ``minihalos''.
TVS04 achieved $\tau _{es} = 0.12$ with $f _*$ = 0.05, \fesc\ = 0.05,
and an IMF with 10 -- 100 \msun, but did not vary $f_*$ and $f_{esc}$.
Haiman \& Holder (2003) combined the three factors into a single
efficiency parameter, $\epsilon_{halo} = f_* \gamma _0 f_{esc}$,
and varied this to achieve reasonable fits to the {\it WMAP} and
Gunn-Peterson data. Cen (2003) studied the feedback effects of X-ray
emission from early SN remnants and black hole accretion and found
that low-mass halos could form stars efficiently enough to reionize the
universe completely by $z = 17$.  Though they differ in their details,
the theoretical studies cited above generally agree that dark matter
``minihalos'' (with virial temperatures $T = 10^3 - 10^4$ K, or $M =
10^6 - 10^8$ \msun\ at $z = 6 - 15$), which could probably cool only by
molecular hydrogen (H$_2$), are essential to explaining the {\it WMAP}
data on $\tau _{es}$.  The reasons are the cumulative nature of $\tau
_{es}$ and the bottom-up nature of structure formation.  If the universe
is completely ionized only at $z < 10$, $\tau _{es} \approx 0.08$.
High levels of ionization at $z > 10$ are therefore necessary to match
the {\it WMAP} constraint, which corresponds to complete ionization
to $z = 17$.  In the range $z = 10 - 17$, minihalo hold most of the
collapsed baryons in the $\Lambda$CDM cosmology.  At $z = 15$, 3.3\%
of all baryons reside in collapsed halos with $T_{vir} \geq 10^3$ K,
but only 0.7\% are in objects with $T_{vir} \ge 10^4$ K. If $\sim 80$\%
of the baryons do not participate in star formation, extreme assumptions
about $\gamma _0$ and \fesc\ would be necessary for $\tau _{es}$ =
0.17 or the models may fail altogether.

Dark-matter ``minihalos'' apparently need to form stars with efficiency
$f_* \gtrsim 0.01$ to match the key data on reionization.  However,
these objects have the most uncertain physics of gas cooling and star
formation and are likely to be too faint to be constrained by independent
observation at high redshift. Some hydrodynamical simulations (Abel,
Bryan, \& Norman 2000, 2002; Bromm, Coppi, \& Larson 1999, 2002) have
followed H$_2$ cooling in these objects and determined that they likely
form only very massive stars (VMSs; $M \gg 100$ \msun) with low efficiency
($f_* \sim 0.001 - 0.01$) and then rapidly dissociate the H$_2$ in their
neighboring halos, thereby suppressing further the average minihalo
efficiency. Other authors (Ricotti et al. 2002; Cen 2003) have argued
that the feedback of soft-UV and X-ray photons is likely to be positive,
such that minihalos can experience significant star formation over long
periods.  In light of all these results, efficient star formation in
these halos can be taken as a {\it prediction} of successful reionization
models, most of which would fail without the participation of minihalos.

This study extends previous work (particularly by Ricotti \& Gnedin
2005, hereafter RG05) to argue that these predictions can be tested
already using dwarf galaxies in the Local Group (LG), and that these
dwarf galaxies likely formed as ``minihalos'' at high redshift and
contributed significantly to the reionization of the IGM.  Section 2
of this {\em Letter} briefly describes the framework of Tumlinson
(2006), concentrating on the modifications needed to study LG dwarfs.
Section 3 reviews the data on Local Group dwarfs and presents models for
their luminosity, metallicity, and star formation efficiency.  Section 4
summarizes the conclusions.

\section{Methods} 

Tumlinson (2006, hereafter Paper I) presented a new chemical evolution
code that works both within the hierarchical context of galaxy formation
and in the stochastic limit of low-metallicity Galactic evolution.
Paper I focused on primordial IMF constraints using Galactic halo stars,
but the framework is adaptable to model halos at arbitrary mass and
redshift and so is well suited to modeling pre-Galactic halos that form
before reionization.

The framework uses the common technique of halo merger trees (Somerville
\& Kollatt 1999) to decompose the Galaxy into its precursor halos
working backward in time. It then calculates the history of star
formation and chemical enrichment in these objects working forward in
time, keeping track of all individual metal-producing supernovae and
assigning metallicity to new star formation stochastically from all prior
generations. Stars are formed with a constant efficiency per timestep,
$\epsilon_*$, such that the mass formed into stars $M_* = \epsilon_*
M_{gas} \Delta t$ in time interval $\Delta t$.  At metallicities below
the critical value for normal star formation, $Z_{crit} = 10^{-3.5}
Z_{\odot}$ (Schneider et al. 2002; Santoro \& Shull 2006), the IMF is
a log-normal function with characteristic mass $m_c$ and width $\sigma
_{IMF}$.  Above $Z_{crit}$, a power-law IMF with Salpeter slope ($\alpha
= -2.35$) is assumed, with $0.5 - 140$ \msun.  IMFs with higher $m_c$
or steeper $\alpha$ are more ``top-heavy'', while $Z_{crit}$ effectively
sets the time over which the top-heavy IMF persists.  In the small halos
of interest, metallicity is correlated with time but does not increase
monotonically over short times.

The Paper I method is changed slightly to model dwarf galaxies in
the Local Group.  The model dwarf galaxies are like their real-world
counterparts in that they virialized early, formed stars for a short
period, and then lost their gas, were ionized by internal or external
sources, or otherwise had their star formation truncated at some
redshift, $z_{end}$, the redshift of the root halo in the tree. After
this redshift they evolve no further.  In the real universe, these dwarf
galaxies may continue to evolve dynamically by accretion into larger
objects, tidal stripping, or they may remain in isolation, but they do
not appear to evolve chemically beyond the loss of their gas. A large
grid of models was produced with $M_h = 10^7 - 10^9$ \msun, $\epsilon_* =
0.1 - 3.0 \times 10^{-10}$ yr$^{-1}$, and $z_{end} = 6 - 16$. The output
quantities of interest are sensitive to these quantities but not to the
choice of IMF below $Z_{crit}$ or to the interstellar mixing rates.

\section{Local Group Dwarfs and Reionization} 

The roughly 40 known Local Group dwarf galaxies span a range of types
(from spheroidal to elliptical to irregular), have generally low
metallicity, and dynamical mass from $10^6 - 10^9$ \msun\ (Mateo 1998).
Prompted by the old age of many dSph populations, and by the discrepancy
between CDM predictions for dwarf galaxy numbers and the known population,
recent theoretical studies have posited a close connection between
reionization and the early evolution of LG dwarf galaxies.  Bullock,
Kravtsov, \& Weinberg (2001) suggested that the suppression of star
formation in small halos by ionizing photons at $z > 6$ could explain
the relatively small number of LG dwarfs compared with the CDM prediction.

RG05 examined this ``reionization hypothesis'' in hydrodynamical
simulations that included H$_2$ cooling and radiative feedback.  They were
able to reproduce many of the observed properties of the LG dwarfs in
simulations that stopped at $z = 8.3$, with no further evolution in the
galaxies of interest. They proposed a classification scheme in which LG
dwarfs are divided into three categories: ``survivors'' massive enough to
retain their gas and continue forming stars robustly after reionization,
``polluted fossils'' that were affected by reionization but managed
to continue forming stars at a modest level, and ``true fossils'',
which formed most of their stars before reionization and essentially
none afterward.  I adopt this scheme here and use the ``true fossils'',
those galaxies with exclusively old stellar populations, roughly 1\%
solar metallicity, and $M_V \gtrsim -12$, to constrain the star formation
efficiency of small halos at high redshift and assess the star formation
efficiency of minihalos.

\begin{figure}
\centerline{\epsfxsize=\hsize{\epsfbox{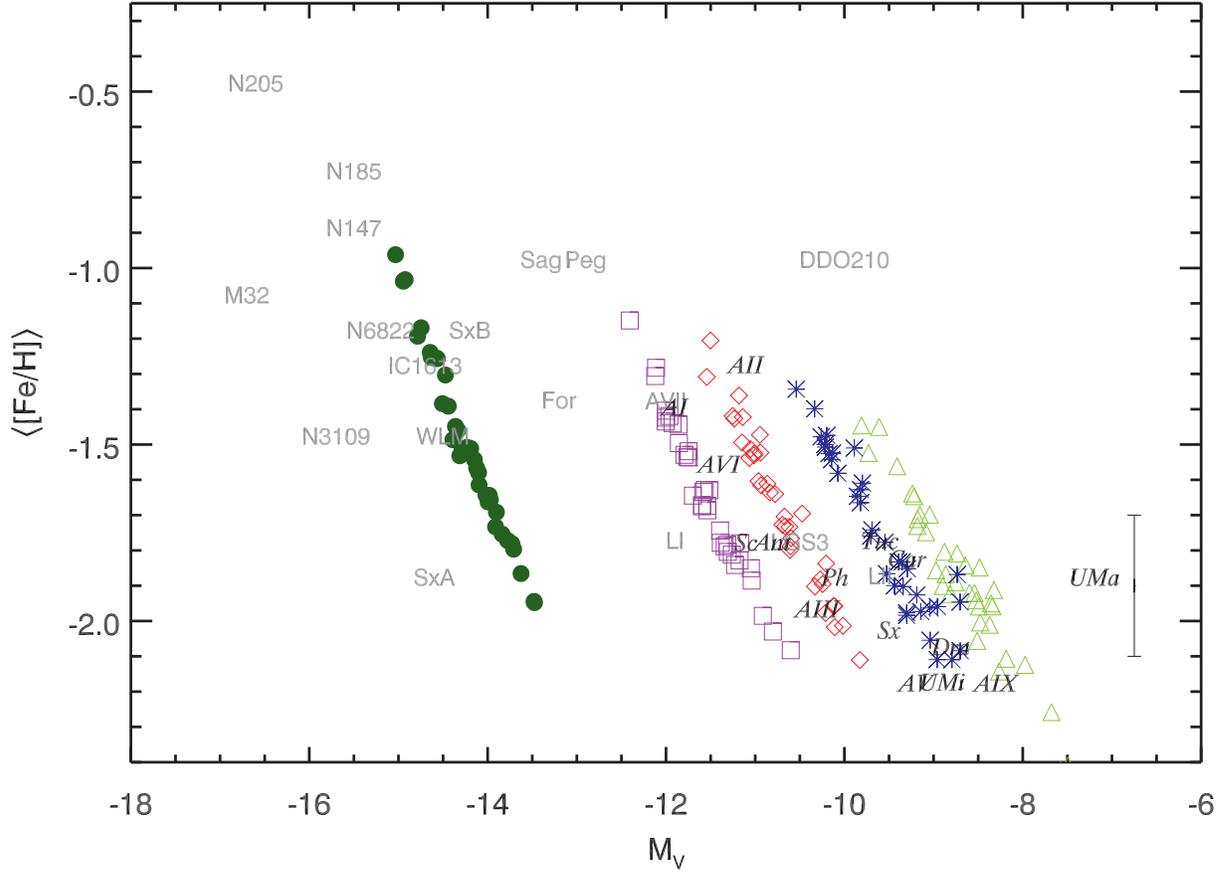}}}
\figcaption{Luminosity-metallicity ($L-Z$) relation for model dwarf
galaxies on constant-mass sequences with $M_h = 10^9, 10^8, 5 \times
10^7, 2 \times 10^7, 10^7$ \msun\ from left to right.  The star formation
parameter $\epsilon$ ranges from $1 - 3 \times 10^{-10}$ yr$^{-1}$ from
the bottom to top of the constant-mass sequences. The Ursa Major dwarf
(Willman et al.~2005) has only a metallicity range. \label{mv-feh-fig}}
 \end{figure}

The first result of this study is that the observed luminosity-metallicity
($L-Z$) relation of LG dwarfs is reproduced exactly with minimal tuning
of the models.  This relation appears in Figure~\ref{mv-feh-fig}, where
a selected group of LG dwarfs is plotted in the style of Caldwell (1999).
Survivors and polluted fossils are marked in gray type; true fossils
are italicized.  Mean metallicities are taken from Mateo (1998),
except for And V (Zucker et al. 2004) and Ursa Major (Willman et
al. 2005). Absolute V-band magnitudes for $z = 0$ are obtained by
interpolation of the tabulated isochrones for 1/50 and 1/200 solar
metallicity from Girardi et al. (2002) and applied individually to 
stars that reside in the final halo at $z_{end}$.  The points mark
the theoretical calculations for $M_h = 10^9, 10^8, 5 \times 10^7,
2 \times 10^7$, and $10^7$ \msun, from left to right; the models are not
intended to perfectly match the survivor and polluted fossils that have
experiences recent star formation (in gray).  Within each constant-mass
sequence, the time-averaged star formation parameter $\epsilon _* = 1 -
3 \times 10^{-10}$ and $z_{end} = 6 - 10$. An increase in $\epsilon_* $ or
decrease in $z_{end}$ moves points up and to the left in the constant-mass
sequences. For the same $\epsilon_{*}$ and $z_{end}$ combination, the
variation in mass traces out a line of the same slope as the observed
trend. We can therefore interpret the observed $L-Z$ relation as a
variation in mass with scatter introduced by small variations in the
star formation histories of individual objects.  This concordance was
achieved with minimal tuning; it was necessary to vary only $\epsilon _*$
with the same mean and in the same range found in Paper I to match
the Galactic halo metallicity distribution.  I conclude that the simple
description of LG dwarf satellites as fossils from before reionization
is accurate and that a more complicated model is unnecessary.

\begin{figure}
\centerline{\epsfxsize=\hsize{\epsfbox{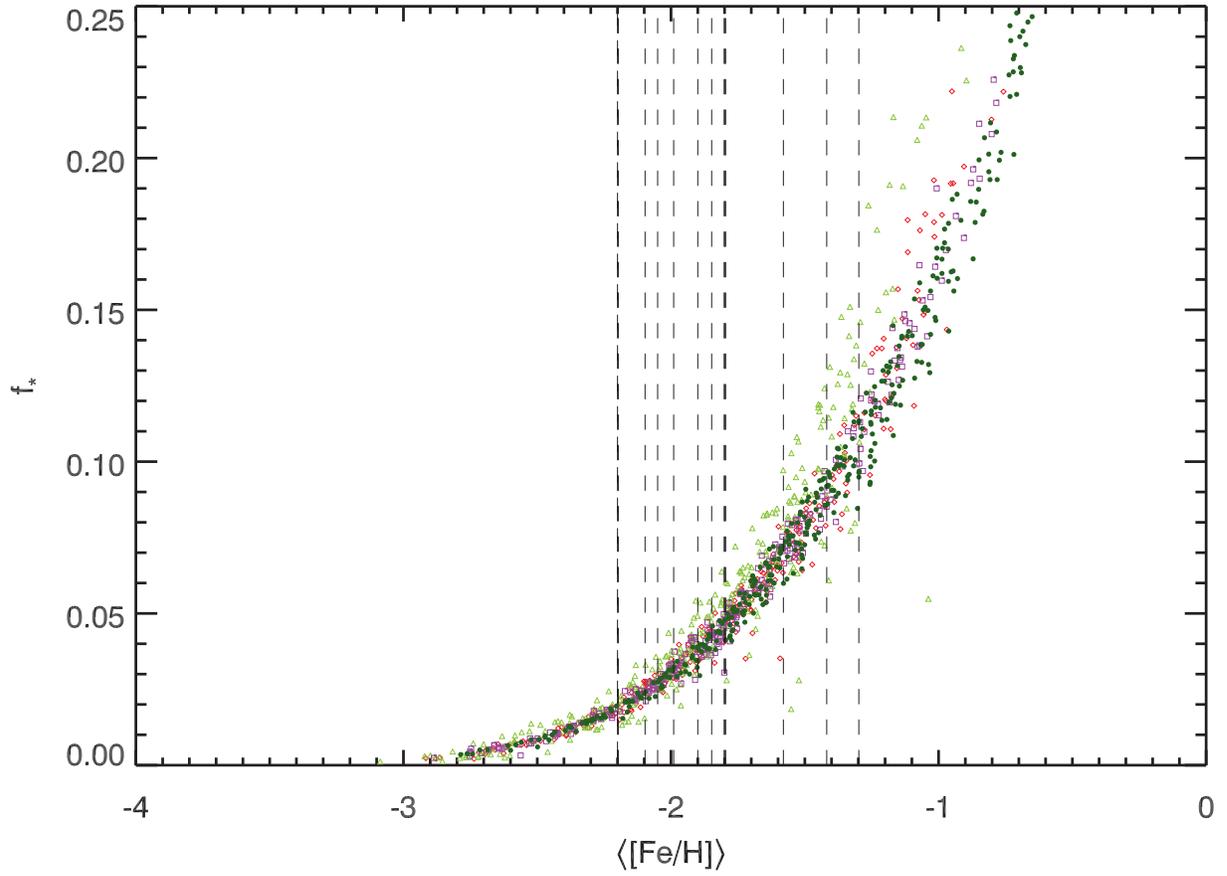}}}
\figcaption{The total star formation efficiency, $f_*$, in small halos
of varying mass that cease forming stars at $z_{end} = 6$. The symbols
are the same as in Figure~\ref{mv-feh-fig}. The correlation of $f_*$
with mean metallicity is strong and insensitive to $M_h$ and $z_{end}$.
\label{fstar-fig}} \end{figure}

The second result of this study is the strong correlation between
total star formation efficiency, $f_{*}$, and mean metallicity
of the resulting dwarf galaxy for thousands of model halos
(Figure~\ref{fstar-fig}). Because metallicity and efficiency are
effectively mass-normalized quantities, this correlation is insensitive
to halo mass. Only at the lowest mass plotted here ($M_h = 10^7$ \msun)
does the relation depart to slightly higher efficiency for a given
metallicity, with increased scatter. This relation can now be used to
determine the star formation efficiency of Local Group ``fossil'' dwarfs.

\begin{figure}
\centerline{\epsfxsize=\hsize{\epsfbox{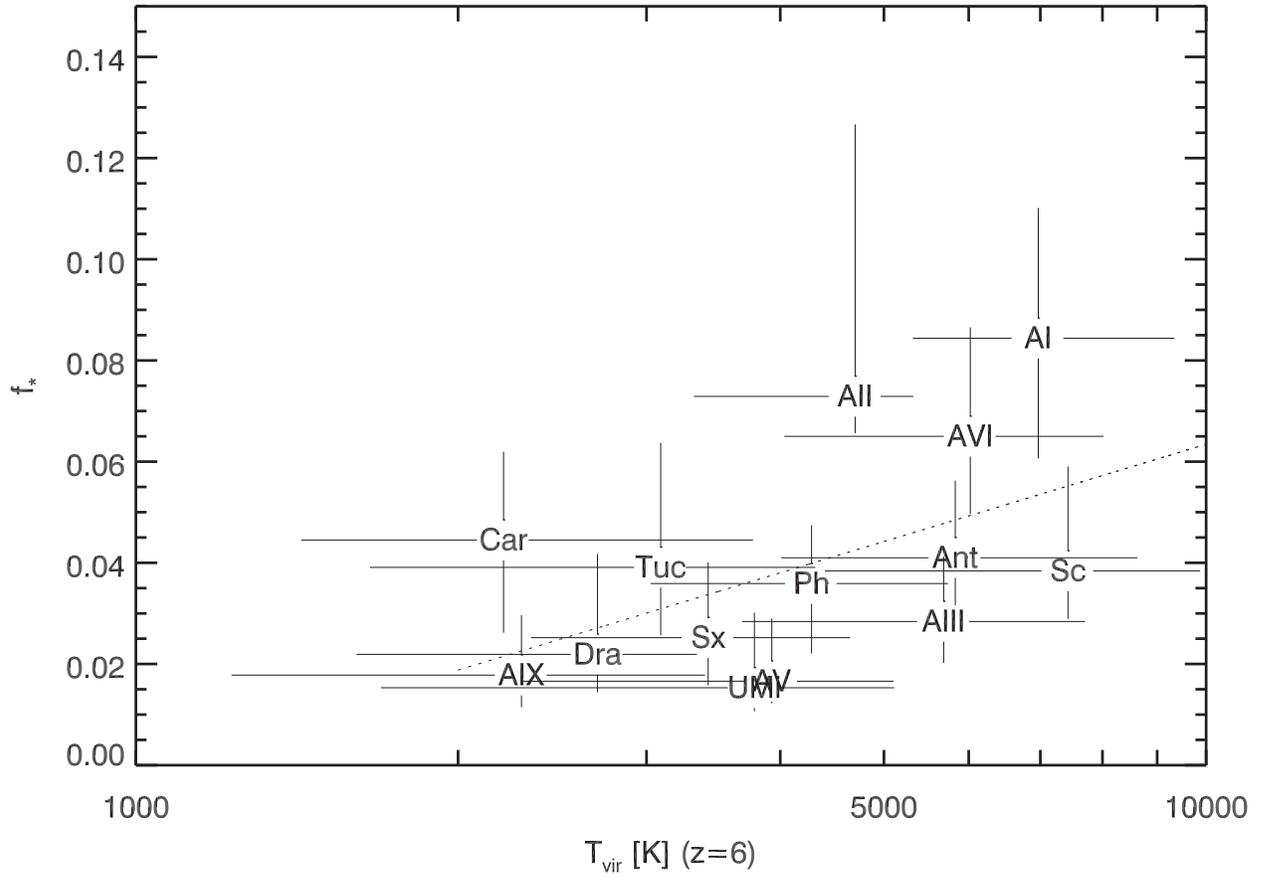}}}
\figcaption{The total star formation efficiency $f_*$ for the LG true
fossils, derived from maximum likelihood fits to the observed mean
metallicity and luminosity.  $T_{vir}$ is evaluated at $z = 6$ using
the best-fitting halo mass, $M_h$, and assuming this is unchanged since
$z = 6$. The error bars mark the $2\sigma$ (95\%) confidence intervals
derived from Markov chain searches of $(M_h, \epsilon _*)$ parameter
space. The dotted line marks a tentative linear fit to $(log T_{vir},
f_*)$.\label{vir-fig}} \end{figure}

Estimates for $f_*$ in the LG fossil dwarfs are obtained by varying the
assumed total halo mass, $M_h$ and time averaged star formation parameter,
$\epsilon _*$, to obtain best fits to the observed $M_V$ and $\langle
[Fe/H] \rangle$, assuming the errors in these quantities tabulated
by Mateo (1998).  The optimization is done with a Monte Carlo Markov
chain maximum likelihood estimator so that confidence intervals are also
obtained.  Figure~\ref{vir-fig} shows the results for total star formation
efficiency $f_*$ compared with $T_{vir}$ derived from fits to the observed
L-Z relation.  For the model ``true fossils'', $f_*$ ranges from $2 -
8$\%, while the best-fitting halo masses clearly indicate that these
galaxies lie in the regime of cosmological ``minihalos'' at high redshift.
There is also tentative evidence for a positive correlation of $f_*$
with $T_{vir}$, at 90\% significance; a linear best fit is shown in the
dotted line.  These results assume a fixed $z_{end} = 6$, but there are no
substantial changes to the individual values or trend if $z_{end} = 10$.
The width of the model metallicity distribution function (MDF) does vary
with $z_{end}$; shorter total star formation times (higher $z_{end}$)
yield narrower distributions.  Halos that form stars until $z = 6$
have broader distributions than those that had their star formation
truncated earlier, but the existing observations of $\sigma([Fe/H])$
are too imprecise to use this as a sensitive indicator of the duration of
their bursts.  Constraints on models of the detailed star formation and
feedback mechanisms in these objects would benefit from the more sensitive
and precise observations of MDF and relative chemical abundances that
future large telescopes could provide.

Based on these results I conclude that minihalos contribute significantly
to the reionization of the IGM at $z > 6$. However, we cannot be sure
that the minihalos we see today as fossil dwarf galaxies in the LG
are typical of minihalos that formed in this region of space,
or that the LG population is representative of the minihalo population
as a whole. Such a concern could be answered by detailed examination
of dwarf galaxies outside the Local Group, but this task is probably
too challenging for observations in the near future. Future theoretical
work should attempt a full chemo-dynamical treatment of the Local Group
to help assess the bias of the LG dwarfs in mass or metallicity,
which may change their place in the scheme of reionization.

\begin{figure}
\centerline{\epsfxsize=\hsize{\epsfbox{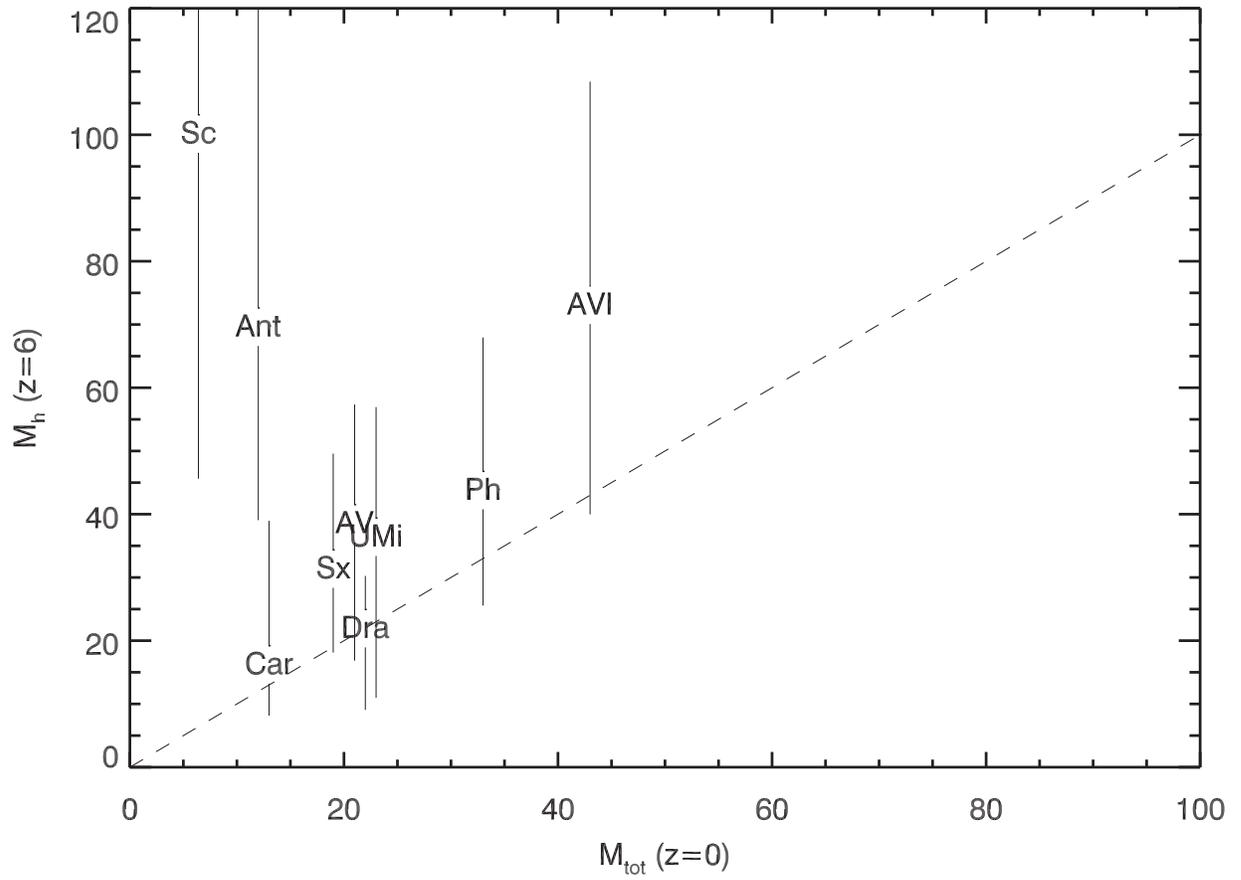}}}
\figcaption{Comparison of total halo mass, $M_h$, inferred from halo
merger tree models with dynamical masses, $M_{tot}$. Error bars show
2$\sigma$ (95\%) confidence intervals.  \label{mcom-fig}}
\end{figure}

Total masses inferred from the L-Z relation in the process of
constructing Figure~\ref{vir-fig} are compared to their dynamically
estimated masses in Figure~\ref{mcom-fig}, where the best-fit $M_h$
is plotted against the measured $M_{tot}$ for the nine fossils with
rotation-curve mass estimates (Mateo 1998).  The models generally best
match the observed L-Z relation for masses similar to the dynamical
estimates (a 1:1 ratio is marked with the dashed line).  Thus it does
not seem necessary to invoke significant mass loss from the ``true
fossils'' to explain any of their observed properties, except in a
few cases.  By this measure, Sculptor, Antlia, and And VI appear to
have lost significant mass in this interval. This apparent mass loss,
compared with other galaxies, can be taken as a prediction of the models
presented here that may be testable by more detailed examination of
their stellar populations.

\begin{figure}
\centerline{\epsfxsize=\hsize{\epsfbox{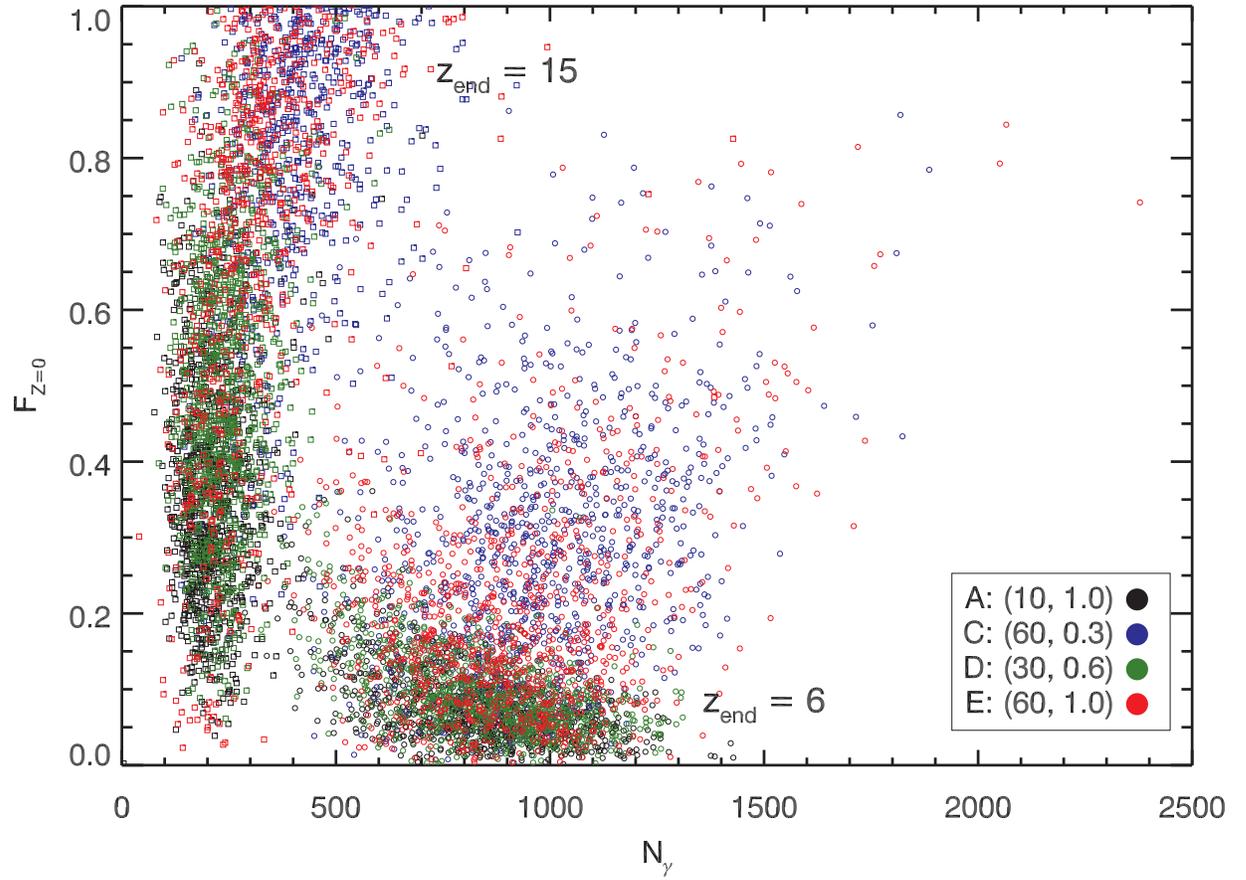}}}
\figcaption{Ionizing photon fraction from metal-free stars, $F_{Z=0}$,
versus total ionizing photon production, $N_\gamma$, for four of the
IMF test cases in Tumlinson (2006), e.g. IMF A has $m_c = 10$ \msun\
and $\sigma_{IMF} = 1.0$.  \label{fg-fig}} \end{figure}

The merger-tree models keep individual account of all supernova-forming
stars, so the distribution of stellar metallicities is known exactly at
all times.  The models specify the fraction of ionizing photons produced
by stars of different metallicities.  Figure~\ref{fg-fig} shows the
fraction of ionizing photons from metal-free stars, $F_{Z=0}$, plotted
against the total ionizing photon budget per baryon in the final halo
($N_\gamma \approx f_* \gamma _0 \gtrsim 1000$ for the $\tau_{es} \gtrsim
0.1$ models of TVS04) for four of the five IMF test cases of Tumlinson
(2006). For $z_{end} = 6$ (open circles), the mean $F_{Z=0}$ is $\approx
5$\% for IMFs A and D. Results for IMF B are similar to case A and are not
shown in the figure. Cases C and E are more top-heavy and give $F_{Z=0}
\approx 0.5$ with a long tail to high $F_{Z=0}$. Paper I showed that
cases A, B, and D, provided the best overall fit to the data on Galactic
chemical evolution, while cases C and E generally overproduce very massive
stars that are inconsistent with the abundance data.  If $z_{end} = 15$
(open squares), the relative contribution of metal-free stars is larger,
$F_{Z=0} \approx 0.1 - 0.5$ for cases A and D, but here stars are formed
for a shorter total time and so have reduced $N_{\gamma} \sim 200$.
As reionization models typically require $N_{\gamma} \gtrsim 1000$ from
minihalos (TVS04, HH03) to achieve $\tau _{es} \gtrsim 0.1$, it seems
that a successful model cannot reproduce the WMAP $\tau _{es}$ and also
have metal-free stars as the primarily agents of reionization.  If the
minihalos that survive are typical of those that formed during the epoch
of reionization, it appears that metal-free stars started but could not
have completed this critical transition in the state of matter in the IGM.

\section{Conclusions} 

Based on halo-merger-tree chemical evolution models of Local Group 
``fossil'' dwarfs, I conclude that: 
\begin{itemize} 
\item[1.] Models of small dark matter halos forming stars during
        reionization accurately reproduce the observed
        luminosity-metallicity relation for LG dwarf galaxies with
        the same model parameters that also match the 
        Galactic stellar halo.
\item[2.] The strong correlation between mean metallicity
        and total star formation efficiency in ``true fossils'' implies
        total star formation efficiency $f_* = 2 - 8$\% for halos
        with mass $M_h = 10^7 - 10^8$ \msun, or $T_{vir} = 2000 - 7000$
        K at $z = 6 - 15$.
\item[4.] By tracking chemical enrichment inside these small halos,
        the models also specify that metal-free stars contribute
        approximately $5-10$ \% of the total ionizing photon budget
        from minihalos to $z = 6$. Thus metal-free stars probably did
        not provide the majority of ionizing photons to the IGM.
\end{itemize}

I conclude that dwarf galaxies that form all their star early and
survive into the present-day Local Group provide sensitive indicators
of the star formation history of low-mass ``minihalos'' during the
epoch of reionization. By comparison with the results of RG05, which
stimulated this work, my results more accurately describe the present-day
luminosity-metallicity relation and independently derive the total
star formation efficiency and first-stars contribution to the ionizing
budget. Conversely, these models lack the sophisticated treatment of
radiative feedback and three-dimensional hydrodynamical evolution that
simulations allow and do not account for possible biases in the surviving
dwarf galaxy populations.  The two approaches therefore complement each
other and show how Local Group fossils provide potentially critical tests
of competing scenarios for reionization.  These results suggest that no
reionization model should be considered successful unless it correctly
incorporates the metallicity and age of old LG stellar populations into
the history of this critical epoch. Because LG dwarf galaxies provide
important information that is not available directly at high redshift,
we may have as much to learn about reionization from $z = 0$, where we
can study it residues, as at $z = 6$, where it ended.

\acknowledgements
Thanks to Beth Willman, Connie Rockosi, Raja Guhathakurta, Kathryn
Johnston, and Arieh Maller for instigating discussions, and to Priya
Natarajan and Paolo Coppi for helpful comments that improved the
presentation of the results. I gratefully acknowledge the generous
support of the Gilbert and Jaylee Mead Fellowship in the Yale Center
for Astronomy and Astrophysics.

\end{document}